\documentclass[lettersize,journal]{IEEEtran}
\usepackage{amsmath,amsfonts}
\usepackage{siunitx}
\usepackage{mathrsfs}
\usepackage{mathtools}
\usepackage{algorithmic}
\usepackage{algorithm}
\usepackage{array}
\usepackage[caption=false,font=normalsize,labelfont=sf,textfont=sf]{subfig}
\usepackage{textcomp}
\usepackage{stfloats}
\usepackage{url}
\usepackage{verbatim}
\usepackage{graphicx}
\usepackage{multirow}
\usepackage{cite}
\usepackage{braket}
\newcommand{\ot}{\otimes}

\newcommand{\bkt}[2]{\langle{#1}|{#2}\rangle}

\begin{document}

\title{Experimental demonstration of response-enhanced fiber Sagnac gyroscopy via imaginary weak-value amplification}
\author{
\IEEEauthorblockN{
Jing-Hui Huang\IEEEauthorrefmark{1},
Hong-Liang Cui\IEEEauthorrefmark{2},
Tianying Chang\IEEEauthorrefmark{2},
Jing Zhang\IEEEauthorrefmark{3},
TianYe Huang\IEEEauthorrefmark{3},
Xiang-Yun Hu\IEEEauthorrefmark{1}
}\\
\IEEEauthorblockA{
\IEEEauthorrefmark{1}Hubei Subsurface Multi-scale Imaging Key Laboratory, 
School of Geophysics and Geomatics, China University of Geosciences, Wuhan 430074, China\\
\IEEEauthorrefmark{2}Shenzhen Institutes of Advanced Technology, Chinese Academy of Sciences, Shenzhen 518000, China\\
\IEEEauthorrefmark{3}School of Future Technology, School of Mechanical Engineering and Electronic Information, China University of Geosciences, Wuhan 430074, China
}
\thanks{Corresponding author: Xiang-Yun Hu (xyhu@cug.edu.cn).}
}


\maketitle

\begin{abstract}
We experimentally demonstrate a 100-m long fiber Sagnac gyroscope employing imaginary weak-value amplification (IWVA) to enhance the optical readout response to rotation. A Sagnac-induced polarization phase is converted into a measurable spectral response through polarization post-selection and spectrally resolved detection. By varying the post-selection angle, we characterize the gyroscope response over the experimentally accessible rotation-rate range. The response increases systematically toward orthogonal post-selection, yielding a maximum experimental enhancement of approximately 13 times relative to a fixed reference configuration without postselection. Fast Fourier transform and Allan-deviation measurements further characterize the noise and stability of the gyroscope. Near-orthogonal post-selection reduces the equivalent rotation-rate fluctuations, with the minimum Allan deviation decreasing from $0.06~\mathrm{rad/s}$ for the reference configuration to $0.0091~\mathrm{rad/s}$ for the configuration exhibiting the maximum response enhancement, corresponding to an approximately 6.6-fold reduction. Low-frequency fluctuations of the broadband superluminescent diode source limit the long-term stability. These results demonstrate that IWVA provides a practical approach for response-enhanced fiber-based Sagnac gyroscopy and precision rotation sensing.
\end{abstract}


\section{INTRODUCTION}

Accurate rotation-rate measurement is essential in various scientific fields, including inertial navigation~\cite{Review-Gyroscope}, rotational seismology~\cite{Review-RotationalSeismometers}, and the detection of Earth's Chandler wobble and annual oscillation~\cite{Schreiber:13,PhysRevLett.107.173904}. 
Sagnac interferometers are primarily implemented as Fiber Optic Gyroscopes (FOGs)~\cite{Culshaw_2006,Jaroszewicz2011,Review-RotationalSeismometers,9989448} and Ring Laser Gyroscopes (RLGs)~\cite{PhysRevLett.107.173904,GINGER-Gyroscopes,Liu:19}. 
FOGs typically employ modulation-demodulation techniques to detect rotation-induced changes in interference intensity~\cite{Jaroszewicz2011}, whereas RLGs measure the resonance-frequency difference between counter-propagating beams. 
%For example, the 16 m$^2$ G-ring achieved a linear drift of 0.162 (prad/s)/day over 60 days~\cite{PhysRevLett.107.173904}. 
Resonant fiber-optic gyroscopes (RFOGs), also known as second-generation FOGs, offer potential advantages in optical stability and miniaturization~\cite{Sanders:21,Zhao:22,review-RFOG}. 
Zhao \textit{et al.} demonstrated a navigation-grade RFOG with a bias instability of 0.009 deg/h and an angle random walk of 0.0093 deg/$\sqrt{h}$~\cite{Zhao:22}. 
Recently, an optical quantum interferometer using maximally path-entangled states achieved a sensitivity of $5 \times 10^{-6}$ rad/s~\cite{doi:10.1126/sciadv.ado0215}.

Applications in navigation increasingly require miniaturized fiber-optic gyroscopes with high sensitivity, reliable stability, and shorter sensing fibers. Because the rotation-induced Sagnac phase scales with the effective area enclosed by the optical path, reducing fiber length and coil dimensions presents a fundamental sensitivity challenge. Thus, improving the signal-to-noise ratio (SNR) within a limited sensing volume is a key objective. Conventional approaches based on increasing the effective area, optimizing optical power, and suppressing technical noise may impose constraints on device size and complexity. Weak-value amplification (WVA)~\cite{RevModPhys.86.307,XU2024100518,10.1063/5.0230512} offers an alternative by tailoring optical preselection and postselection to enhance the measurable response to small rotation-induced phase shifts.

Introduced by Aharonov \textit{et al.}~\cite{AAV}, WVA is a versatile metrological technique for enhancing the readout of weak physical effects~\cite{PhysRevLett.111.033604,doi:10.1063/1.4976312,PhysRevLett.125.080501,Yin2021,PhysRevA.105.013718,PhysRevLett.128.040503,doi:10.1063/5.0132992,Xia2023,PhysRevLett.134.080802,g33l-rgdh,fggq-yhz8,Cao:26}. A weak interaction couples the measured parameter to a meter degree of freedom, and nearly orthogonal postselection can map small parameter-dependent changes onto measurable shifts of the meter. Depending on the pointer and detection scheme, WVA can amplify spatial, temporal, or spectral observables. In particular, imaginary weak values enable small phase shifts to be converted into spectral displacements of broadband optical pointers~\cite{PhysRevLett.111.033604}.

The metrological advantage of WVA is not universal; it depends on the measurement architecture, postselection probability, photon-number constraint, and dominant technical noise~\cite{PhysRevX.4.011031,PhysRevA.94.012329,AWVA,PhysRevA.107.042601}. Its benefits should therefore be assessed under specific experimental constraints rather than inferred solely from the weak-value magnitude~\cite{PhysRevA.109.042602,PhysRevApplied.22.054068}. Time-momentum-correlated weak measurement has also been demonstrated to overcome technical-noise limitations, including jitter and pixelation, and improve precision over conventional weak measurement~\cite{755l-23f6}. WVA may be particularly useful when technical noise, detector saturation, or limited dynamic range constrains conventional readout.

These features make WVA attractive for Sagnac-based rotation sensing in compact fiber systems. In FOGs, rotation is typically inferred from optical intensity or phase, while performance may be limited by source intensity noise, polarization fluctuations, backscattering, and other technical noise~\cite{58063,Culshaw_2006,Zhang:19}. WVA offers an alternative readout in which small phase perturbations are converted into amplified observables while retaining the common-path and compact characteristics of fiber implementations~\cite{8067480,Luo:20,Liu:22}. For example, Luo \textit{et al.} demonstrated a WVA-based fiber-optic hydrophone with a pressure sensitivity of $1.3\times10^{-6}$~Pa/$\mathrm{Hz}^{1/2}$ at 10~Hz~\cite{Luo:20}, while Liu \textit{et al.} reported a noise floor approximately two orders of magnitude below that of conventional fiber-optic interferometric sensors~\cite{Liu:22}. These results demonstrate the potential of WVA for weak-signal readout in fiber sensing.

WVA has also been explored in optical gyroscopes. An integrated Sagnac interferometer using inverse WVA was experimentally demonstrated for rotation measurement~\cite{Song2022}. More recently, an integrated gyroscope combining inverse WVA with a ring resonator was proposed, with theoretical analysis predicting improved angular-rotation readout under suitable detector and optical-power conditions~\cite{Yanik:26}. These approaches use inverse WVA and integrated or temporal readout schemes, distinct from the polarization-mediated imaginary-WVA approach investigated here.

Here, we experimentally investigate imaginary WVA in a polarization-maintaining (PM) fiber Sagnac gyroscope. The Sagnac phase is encoded in polarization and converted into a measurable spectral response through polarization postselection and spectrally resolved detection. By varying the postselection angle, we characterize the measured response coefficient $k(\beta)=\partial\delta I/\partial\Omega$ and quantify relative response enhancement without assigning an absolute weak value to the detector signals. Near-orthogonal postselection, following the WVA approach of Ref.~\cite{PhysRevLett.111.033604}, yields a maximum experimental response enhancement by a factor of 13 relative to a fixed reference configuration. We further characterize the noise spectrum and long-term stability using Fourier analysis and Allan deviation, identifying the main limitations of the present implementation. These results provide an experimental basis for optimizing polarization postselection and spectral readout in compact fiber-based rotation sensing.

\section{THEORETICAL FRAMEWORK}

\subsection{Sagnac phase in optical fiber setup}

The principle of rotation rate measurements is based on the Sagnac effect~\cite{1913L}. 
The optical minimum configuration architecture involves directing light from a source through a beam splitter, resulting in two beams circulating clockwise and counterclockwise within a loop. 
Consequently, the different round-trip times induce the Sagnac phase shift $ \phi_{s}$ when the loop rotates at rate $\Omega$, which can be calculated by:
\begin{eqnarray}
\label{Eq:Sagnac-phase}
 \phi_{s}=\frac{4 \pi R L}{\lambda_{0} c} \times { {\Omega}} ,
\end{eqnarray}
where $R$ and $L$ represent the radius and length of the sensing fiber coil, respectively. 

\begin{figure}[t]
\centering
\includegraphics[width=0.5\textwidth]{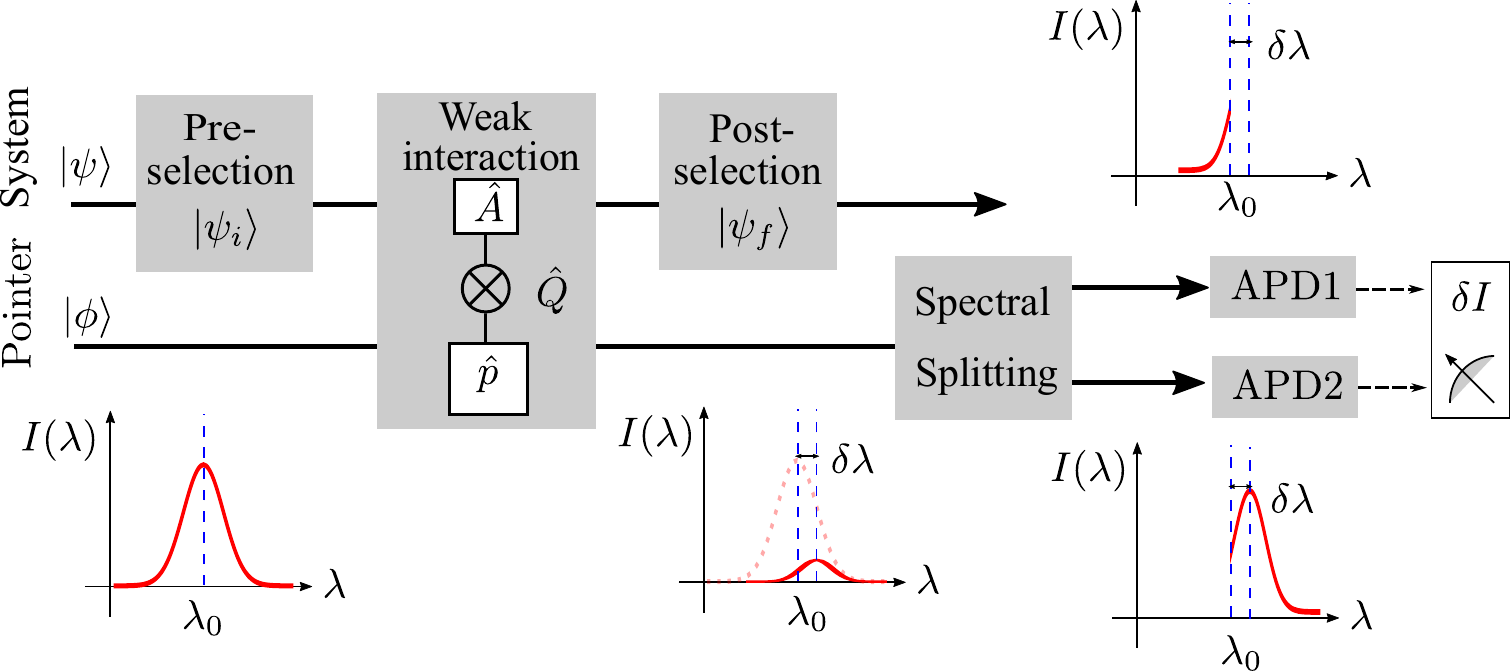}
\caption{\label{Fig:TheoryAndSimulation} 
Principle of imaginary weak-value amplification. A small relative phase is converted into a spectral shift of the optical pointer through polarization post-selection, enabling sensitive frequency-domain measurements.
}
\end{figure}
\subsection{Imaginary weak-value amplification}
We implement a standard weak-value amplification involving imaginary weak values (IWVA)~\cite{PhysRevLett.111.033604} in a PM fiber Sagnac gyroscope. 
As shown in Fig.~\ref{Fig:TheoryAndSimulation},
IWVA is implemented by introducing a weak interaction between a two-level system $ \ket{\psi_{}}$ and a pointer $ \ket{\phi_{}}$ in the momentum space. The weak interaction is characterized by a unitary evolution $\hat{Q} = {\rm exp}(-i g \hat{A} \ot \hat{p})$, where $\hat{A}$ and $g$ represent the observable quantity and coupling strength. The coupling strength $g= \phi_{s} /p_{0}$ is related to the Sagnac phase $\phi_{s}$, corresponding to the rotation rate $\Omega$. 
Here, $p$ is the conjugate momentum. 
%Both the center value of the momentum $p_0$ as well as the full distribution of $p$ can be measured using a spectrometer.
Normally, the qualities of interest can be encoded as the weak coupling strength and can be amplified by a factor known as the weak value:
\begin{equation}
\label{Eq:defination_weakValue_first}
A_{w}:=\frac{ \langle \psi_{^{f}}| \hat{A}\ket{\psi_{i}}}{\bkt{\psi_{^{f}}}{\psi_{i}}},
\end{equation}
where $ \ket{\psi_{i}}$ and $ \ket{\psi_{f}}$ represent pre- and post-selected states of the system, respectively.
%To obtain a weak value with an imaginary part, the two states $ \ket{\psi_{i}}$ and $ \ket{\psi_{f}}$ need to be carefully prepared.
Following the commonly employed IWVA scheme with white sources in free space~\cite{PhysRevLett.111.033604}, the pre-selected state is prepared as a linearly polarized state
\begin{equation}
\label{Eq:pre-selection}
\ket{\psi_{i}}=\sin{ \left( \frac{\pi}{4} \right) } \ket{H} + \cos{ \left( \frac{\pi}{4} \right)} \ket{V},
\end{equation}
where  $\ket{H}$ and $ \ket{V}$ represent the orthogonal polarization bases, respectively. Then, we project the system state into
\begin{equation}
\label{Eq:post-selection1}
| {\psi_{f0}} \rangle = \sin{\left( -\frac{\pi}{4} +\beta \right)} \ket{H} 
+  \cos{\left( -\frac{\pi}{4} +\beta \right)} \ket{V},
\end{equation}
Note that the imaginary part of the weak value emerges from the elliptically polarised system states. 
In this work, by employing a PM Sagnac fiber coil, three factors influence the post-selected state $| {\psi_{f0}} \rangle$ and introduce a phase difference
\begin{equation}
\label{Eq:post-selection1}
\phi_{} = \phi_{s}+\phi_{b} +\phi_{p},
\end{equation}
between $\ket{H}$ and $\ket{V}$. 
Firstly, the weak interaction automatically generates the phase difference, i.e., the Sagnac phase $ \phi_{s}$. The same mechanism is described in Ref.~\cite{PhysRevLett.111.033604}. 
Secondly, photons propagating along the PM fibers induce a phase difference $ \phi_{b}=2 \pi \Delta n L/ \lambda$ due to the birefringence effect, where $L$ is the length of the fiber and $\Delta n $ represents the birefringence coefficient. 
Thirdly, we introduce a compensation phase $ \phi_{p}$ in the experiment to cancel the birefringence $ \phi_{b}$, i.e., $ \phi_{p}+ \phi_{b}=0$, achieved through a polarization controller. 
Consequently, the final post-selected state we prepared is as follows:
\begin{equation}
\label{Eq:post-selection3}
| {\psi_{f}} \rangle = e^{-i\phi_{}/2} \sin{\left( -\frac{\pi}{4} +\beta \right)} \ket{H} 
+ e^{i\phi_{}/2} \cos{\left( -\frac{\pi}{4} +\beta \right)} \ket{V}
\end{equation}
By employing the observable operator $\hat{A}=\ket{H} \bra{H}-\ket{V} \bra{V}$, we can obtain the imaginary part of the  weak value
\begin{equation}
\label{Eq:Im_Aw}
\begin{aligned}
\operatorname{Im} A_w
&= \operatorname{Im} \left\{ 
     \frac{ \langle \psi_f | \hat{A} | \psi_i \rangle }
          { \langle \psi_f | \psi_i \rangle } 
   \right\} \\
&= \operatorname{Im} \left\{ 
     \frac{ 1 - e^{i\phi} \cot\left(-\frac{\pi}{4} + \beta\right) }
          { 1 + e^{i\phi} \cot\left(-\frac{\pi}{4} + \beta\right) } 
   \right\} \\
&= \frac{ \sin\phi \cos 2\beta }{ 1 - \cos\phi \cos 2\beta },
\end{aligned}
\end{equation}

\begin{figure}[t]
\centering
\includegraphics[width=0.5\textwidth]{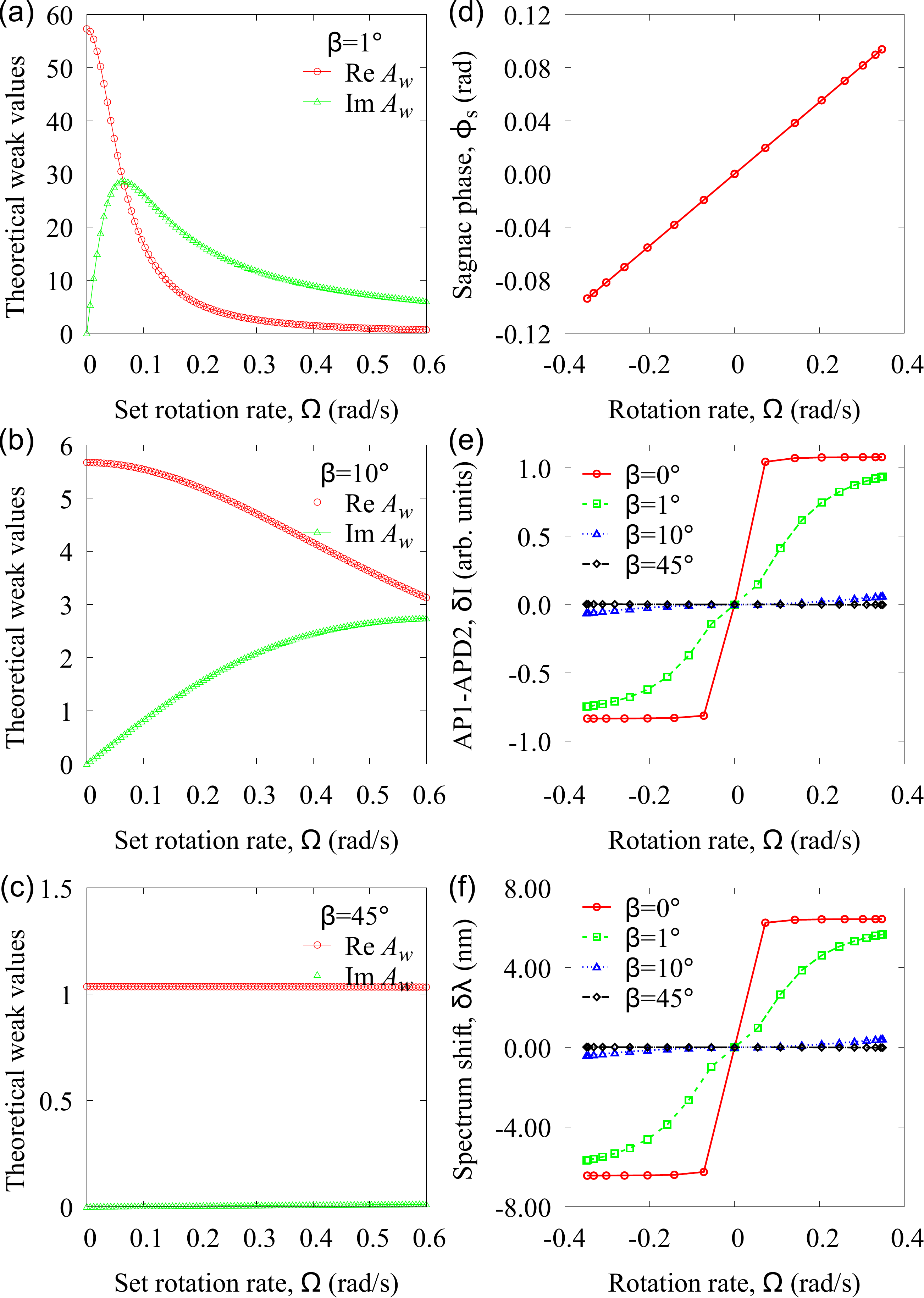}
\caption{\label{fig:simulation}
Simulation results for the sensitivity-enhanced fiber Sagnac gyroscope. The simulation parameters are $R=0.1$ m, $L=100$ m, $\lambda_0=1550$ nm, and $\Delta\lambda=50$ nm. (a--c) Theoretical weak values for the WVA configuration with post-selection angles $\beta=1^\circ$, $10^\circ$, and $45^\circ$, respectively, showing the dependence of ${\rm Re}A_w$ and ${\rm Im}A_w$ on the Sagnac phase. (d) Sagnac phase $\phi_s$, (e) spectral shift $\delta\lambda$, and (f) power difference $\delta I$ as functions of the rotation rate $\Omega$.
}
\end{figure}

In the imaginary-WVA scheme~\cite{PhysRevLett.105.010405,PhysRevLett.111.033604}, a small relative phase $\phi$ between the two-level states can be mapped onto a pointer shift in the momentum domain,
\begin{equation}
\label{Eq:momentumShift}
\delta p = 2 \frac{\phi}{p_0}(\Delta p)^2 \operatorname{Im}A_w .
\end{equation}
For an optical spectral pointer, using $\hbar=1$ and the relation $p=2\pi/\lambda$, and assuming $\Delta p\ll p_0$, we have $\Delta p\simeq 2\pi\Delta\lambda/\lambda_0^2$ and $\delta\lambda\simeq-2\pi\delta p/p_0^2$. The momentum shift can therefore be expressed as a spectral displacement,
\begin{equation}
\label{Eq:spectrumShift}
\delta\lambda
= -\frac{4\pi\phi}{p_0}
\left(\frac{\Delta\lambda}{\lambda_0}\right)^2
\operatorname{Im}A_w
= -\frac{8\pi RL}{c}
\left(\frac{\Delta\lambda}{\lambda_0}\right)^2
\operatorname{Im}A_w \times \Omega,
\end{equation}
where $\Delta\lambda$ denotes the spectral bandwidth of the optical source, and we have used the Sagnac phase $\phi=\phi_s=4\pi RL\Omega/(\lambda_0 c)$. Thus, the imaginary part of the weak value determines the magnitude of the WVA-induced spectral response to the Sagnac phase. This mechanism provides a frequency-domain readout in which a small rotation-induced polarization phase is converted into a measurable spectral displacement of a broadband optical pointer.

Figure~\ref{fig:simulation} shows the theoretical real and imaginary parts of the weak value for the WVA protocol implemented in this work. The calculation assumes an ideal system in which the total relative phase between the polarization components is given solely by the Sagnac phase, i.e., $\phi_s$. The results illustrate the dependence of ${\rm Re}A_w$ and ${\rm Im}A_w$ on the post-selection angle $\beta$ and the rotation rate $\Omega$.
For the polarization states adopted in this work, the ideal theoretical model gives ${\rm Im}A_w=0$ at $\beta=45^\circ$, whereas near-orthogonal post-selection produces a finite and enhanced imaginary weak-value response. Therefore, the $\beta=45^\circ$ configuration is not used to define or calibrate an absolute weak value. Instead, it is treated as a fixed experimental reference for comparing the measured gyroscope response under different post-selection conditions. This distinction is important because the experimentally measured response at the reference configuration can contain contributions from nonideal effects that are not captured by the ideal weak-value model.

The pre- and post-selected linear polarization states are chosen following previous implementations of imaginary WVA with broadband optical sources~\cite{PhysRevLett.111.033604}. Alternative choices of pre- and post-selected polarization states, including elliptical or circular polarization states, can modify the weak-value response and avoid the vanishing of ${\rm Im}A_w$ at a particular post-selection angle~\cite{PhysRevLett.105.010405}. In a fiber-based implementation, however, precise preparation and maintenance of arbitrary polarization states can be complicated by residual fiber birefringence and the finite compensation accuracy of the polarization controller. These experimental imperfections can introduce additional relative phases and deviations from the nominal polarization states. Accordingly, the theoretical weak-value calculation is used here to describe the underlying WVA mechanism, while the experimentally observed enhancement is evaluated independently from the measured response coefficient $k(\beta)$.

\subsection{Power-difference readout of the spectrum shift}
\begin{figure*}[b]
\centering
\includegraphics[width=0.9\textwidth]{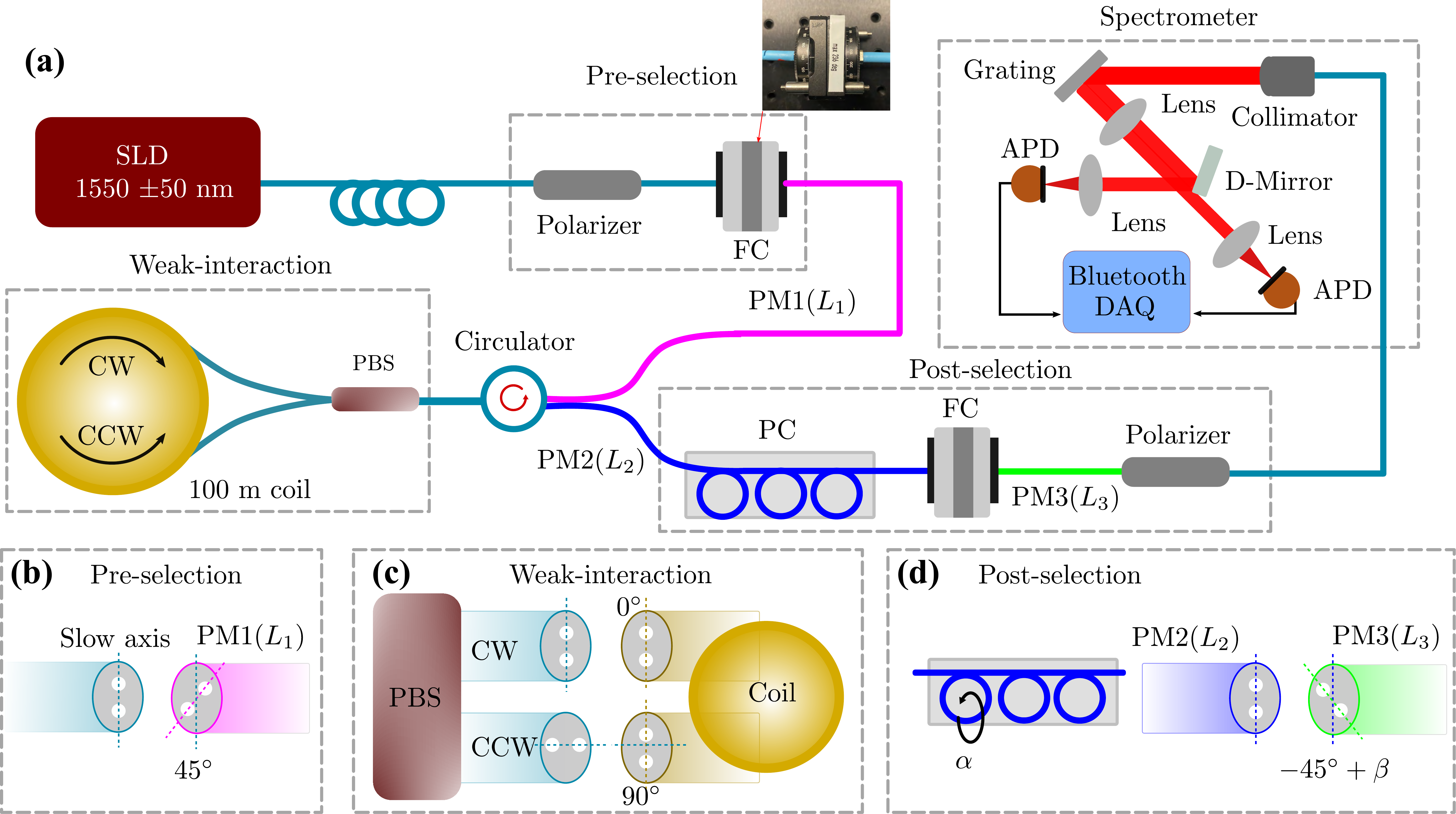} 
\caption{\label{Fig:Experimental_setup} 
(a) Experimental setup.
SLD: Superluminescent Diode;
%Polarizer: in-line fiber polarizer (supplier: THORLABS, ILP1550PM-APC- 1550 $\pm$ 50 nm);
FC: fiber connector;
PM1-PM3: PM fibers with different lengths $l_1$ $l_2$ and $l_3$;
%Circulator: PM Optical Circulator (supplier: Fibermart, 1$\times$2 1550 nm  PM optical circulator both axis working);
PC: polarization controller;
PBS: polarization splitter;
%Collimator: protected silver reflective collimator (supplier: THORLABS, RC12APC-P01);
%Grating: ruled reflective diffraction grating (supplier: THORLABS, GR25-0616, 600/mm, 1.6 $\mu$m blaze);
D-Mirror: Protected Silver D-Shaped Mirror;
APD: Amplified Detector;
%Coil: 100-m PM fiber coil.
Bluetooth DAQ: data acquisition system with Bluetooth communication.
(b), (c) and (d) correspond to the fiber connection angle for pre-selection, weak interaction, and post-selection, respectively.
Two white circles represent panda eyes with the slow axis. 
%Top (e) and side (f) views of the experiment. (g) Photograph of the fiber connector.
}
\end{figure*}

In this work, the spectrum shift is detected using two photodetectors placed at different spectral regions.
%Due to the high cost associated with a high-precision spectrometer for 1550 nm, %and also the difficulty of mounting it on a spinning platform, we opted to construct a simplified version. 
As shown in Fig.~\ref{Fig:TheoryAndSimulation},
spectrum generated by a grating is divided into two parts through a D-mirror, and then the two beams are focused on two intensity detectors. 
The initial position of the D-mirror is adjusted by seeking the power balance between the two detectors when the gyroscope is not rotating.
Finally, we measure the power difference $\delta I$ between the two detectors to estimate the spectrum shift. Mathematically, the $\delta I$ dependence of the spectrum shift $\delta \lambda(\Omega)$ can be derived by assuming a normalized Gaussian-shaped spectrum of the photons:
\begin{equation}
\label{Eq:powerDifference}
\delta I = \frac{\rm APD1-APD2}{ \rm APD1+APD2},
\end{equation}
with
\begin{equation}
\label{Eq:powerDifference1}
{\rm APD1} = \int_{0}^{\lambda_0 - \delta \lambda} {\rm exp} \left( - \frac{(\lambda-\lambda_0- \delta \lambda)^2}{(\Delta \lambda)^2} \right) d\lambda, 
\end{equation}
\begin{equation}
\label{Eq:powerDifference2}
{\rm APD2} = \int_{\lambda_0 - \delta \lambda}^{\infty} {\rm exp} \left( - \frac{(\lambda- \lambda_0- \delta \lambda)^2}{(\Delta \lambda)^2} \right) d\lambda. 
\end{equation}
Note that WVA operates effectively within a specific measurement range characterized by a linear response of $\delta I (\Omega)$. As shown in Figs.~\ref{fig:simulation}(d)-(f), we present simulation results with various post-selections for measuring the rotation rate within the range of $-0.4 \, {\rm rad/s}\leq \Omega \leq 0.4 \, {\rm rad/s}$. Within this measurement range, the anticipated spectrum shifts are considerably smaller than the bandwidth. Additionally, selecting an appropriate post-selection angle ensures a linear response.
When comparing the spectrum shifts in Fig.~\ref{fig:simulation}(e) and the power difference in Fig.~\ref{fig:simulation}(f), 
it is evident that our simplified spectrometer, based on Eq.~(\ref{Eq:powerDifference}), provides a reliable approximation for measuring the spectrum shifts. 
%With this assumption, a linear relationship between the rotation rate $\Omega$ and the detected signal $\delta I $ can be established:
%\begin{equation} \label{Eq:powerDifference-and_rotationRate} \Omega \propto \frac{1}{\rm Im A_w} \delta \lambda \propto \frac{1}{\rm Im A_w} \delta I. \end{equation}
%Since the precise relationship between $\Omega$ and $\delta I$ is challenging to analyze directly, we end up using Eq.~(\ref{Eq:powerDifference-and_rotationRate}) to quantitatively describe the relative magnitude of $\delta I$ when using WVA with different postselection angles.

\section{EXPERIMENT}
The experimental setup of the Sagnac fiber gyroscope based on IWVA is illustrated in Fig.~\ref{Fig:Experimental_setup}. 
For the broadband light source, a 1550 nm superluminescent diode with a near-Gaussian spectral shape was employed (supplier: THORLABS, SLD1005S- 22 mW SLD, 1550 $\pm$ 50 nm). 
To control the PM fiber connection with a manageable ``splicing'' angle, we designed a fiber connector (FC) as shown in Fig.~\ref{Fig:Experimental_setup} by combining a mating sleeve with two cage rotation mounts with a designed 3-D printer component. The sensing fiber, responsible for introducing the Sagnac phase, was made by winding 100 meters of PM fiber with a radius of 0.1 meters (supplier: YOFC, PM 1550-125-18).
%As shown in Fig.~\ref{Fig:Experimental_setup}(f), all the components in Fig.~\ref{Fig:Experimental_setup}(a) are arranged on a rotation breadboard controlled by a high-performance motor (supplier: K-TECH, MG8016-i36).

The WVA process comprised the following four steps.  
(i) ``Pre-selection". We initiated the system state $\ket{\psi_{i}}$ through a polarizer (supplier: THORLABS, FPC030) and a Fiber Connector (FC) set at a $45^{\circ}$ connection angle for two slow axes.
(ii) ``Weak interaction". The light passed through a circulator, 
%(supplier: Fibermart, 1$\times$2 1550 nm  PM optical circulator both axis working),
a polarization combiner/splitter (PBS, supplier: THORLABS, PFC1550A), the Sagnac coil, and then returned through the PBS and circulator. In the coil, establishing a $0^{\circ}$ fiber connection in the clockwise (CW) direction and a $90^{\circ}$ fiber connection in the counterclockwise (CCW) direction ensured that only one axis functioned, thus eliminating the birefringence effect.
(iii) ``Post-selection". A polarization controller (PC), a FC and a polarizer were used to prepare the post-selected state $| {\psi_{f}} \rangle$. In our setup, the birefringence $ \phi_{b}$ mainly originated from the PM1, PM2, and PM3 fibers.
The PC was employed to cancel $ \phi_{b}$ by rotating the first paddle and introducing phase $ \phi_{p}$. We achieved the compensation $ \phi_{p}+ \phi_{s}=0$ by determining the minimum optical intensity after post-selection. 
In addition, the FC and the polarizer, with a $-45^{\circ} + \beta$ cross-connection, provided manageable post-selected states.
(iv) ``Read out" via a simplified spectrometer. A free-space spectrometer was constructed to detect the spectrum shift based on Eq.~(\ref{Eq:powerDifference}).  The collimated beam was incident on a reflective grating (supplier: THORLABS, GR25-0616, 600/mm, 1.6 µm Blaze, 25 $\times$ 25 $\times$ 6 mm). Subsequently, the spectrum was equally divided by a D-Mirror and the two separated spectrums were focused by the lens and detected by detectors (supplier: THORLABS, PDA20CS2). 
The two photocurrents were transmitted to a computer by a data acquisition system with Bluetooth communication (sampling rate 10 Hz, measurement range 0-10 mA, sampling accuracy 0.01 mA).

\section{Results}

\subsection{Sinusoidal rotation measurements}

We first characterized the response of the gyroscope by applying a sinusoidal rotation rate,
\begin{equation}
\label{Eq:AppliedSingal}
\Omega(t)=0.35\times\sin(2\pi f_{\Omega}t)\,\,\mathrm{rad/s},
\end{equation}
where $f_{\Omega}$ denotes the rotation frequency. By varying $f_{\Omega}$, the response of the gyroscope was investigated over different frequencies of the set rotation. Owing to the mechanical load of the optical components and breadboard, the motor cannot reliably generate rotation rates below approximately $0.03$~rad/s. The sinusoidal driving signals were therefore chosen within an experimentally accessible range, with a maximum rotation rate of $0.35$~rad/s and a minimum rotation rate of $0.07$~rad/s.

\begin{figure}[t]
\includegraphics[width=0.5\textwidth]{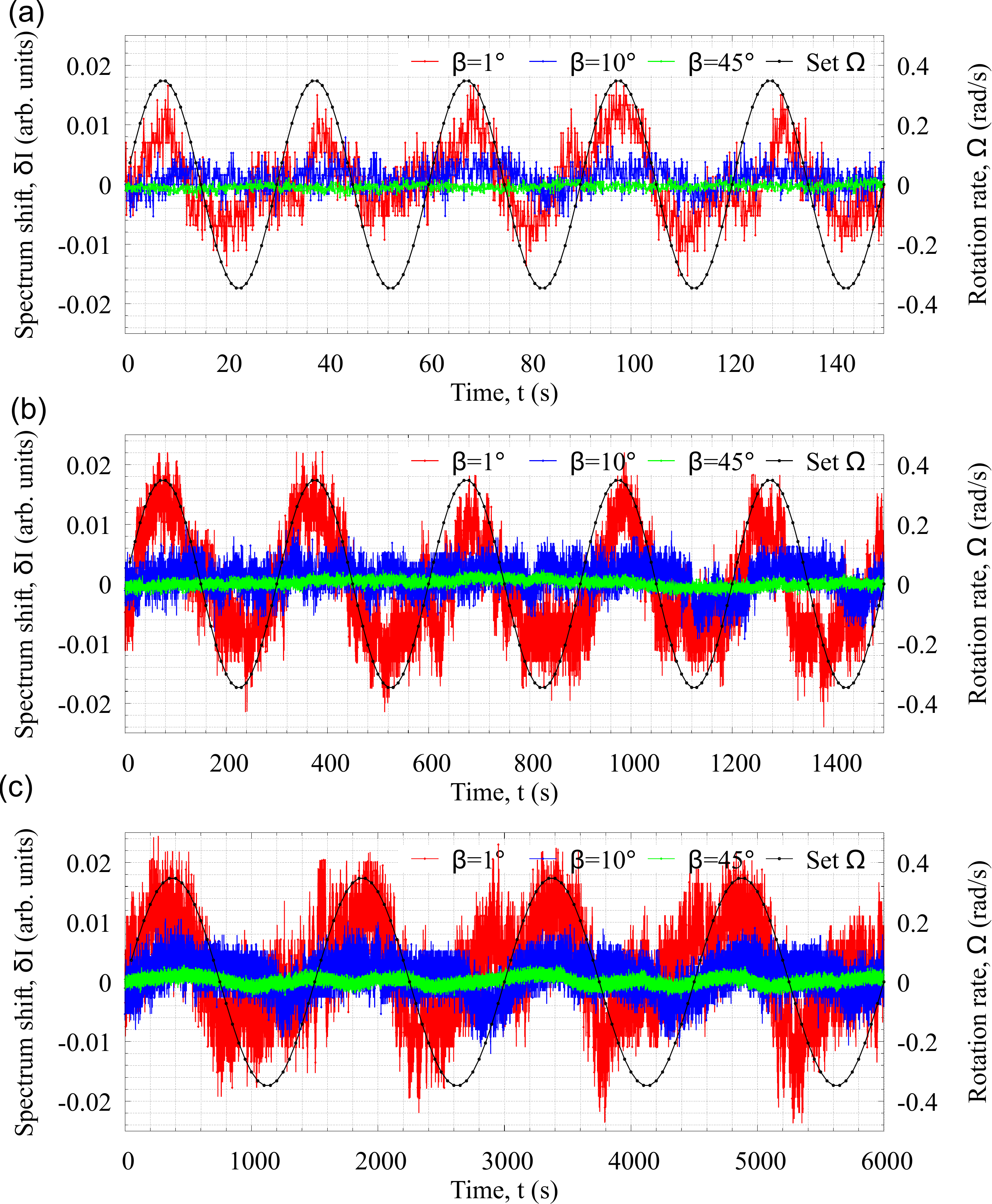}
\caption{\label{Fig:Result_timeDomain}
The response of the spectrum shift $\delta I$ to sinusoidal rotation-rate signals $\Omega(t)=0.35\times\sin(2\pi f_{\Omega}t)$~rad/s. (a)--(c) show the measured detector signals for $f_{\Omega}=0.03$, 0.003, and 0.0006~Hz, respectively.
}
\end{figure}

Figure~\ref{Fig:Result_timeDomain} shows the measured detector signal $\delta I(t)$ for different post-selection angles $\beta$ and rotation frequencies $f_{\Omega}$. For all investigated rotation frequencies, reducing the post-selection angle increases the amplitude of the oscillatory detector response. The rotation signal remains observable at all three frequencies, although the signal-to-background ratio decreases as the rotation frequency is reduced. At the lowest rotation frequency of $0.0006$~Hz, additional low-frequency fluctuations are observed in the measured signal, including a pronounced fluctuation around $t=3000$~s.

These measurements demonstrate that the post-selection angle provides an effective control parameter for the experimentally measured response of the fiber Sagnac gyroscope. The corresponding quantitative response coefficients are determined from the measured dependence of $\delta I$ on $\Omega$ in the following section.

\subsection{Experimental response enhancement}

We next performed steady-rotation measurements to quantify the dependence of the gyroscope response on the post-selection angle. For each $\beta$, the gyroscope was continuously rotated at a constant rate, and the corresponding detector signal $\delta I$ was recorded. The response coefficient was determined from the linear dependence of $\delta I$ on the rotation rate $\Omega$,
\begin{equation}
k(\beta)=\frac{\partial\delta I}{\partial\Omega},
\end{equation}
where the slope was obtained from a linear fit within the experimentally verified linear-response range of $\Omega\leq0.2$~rad/s.

\begin{figure}[t]
\centering
\includegraphics[width=0.5\textwidth]{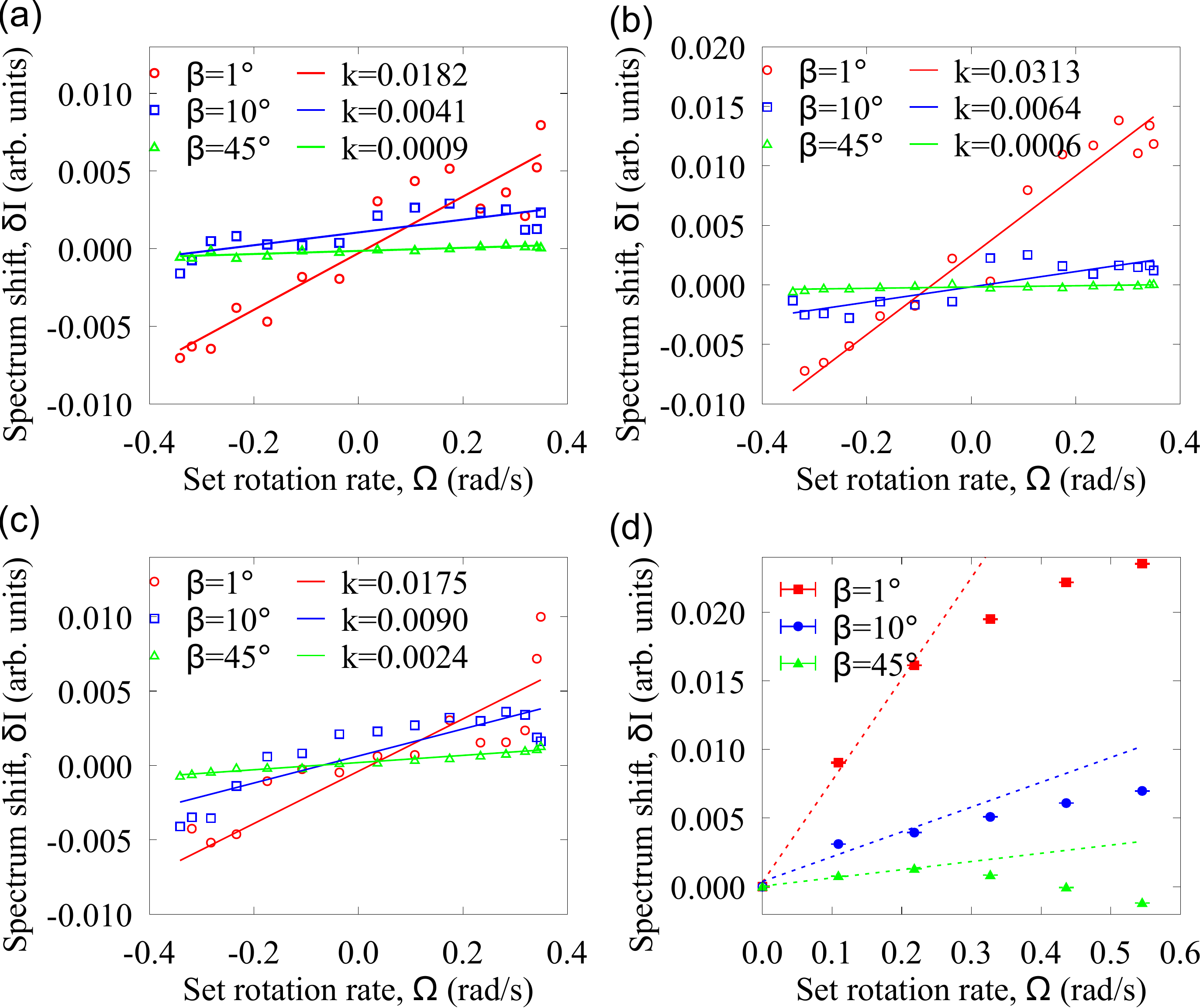}
\caption{\label{Fig:Result_Response_enhancement}
Experimental response enhancement for different post-selection angles.
(a)--(c) Representative linear fits of the detector signal $\delta I$ as a function of rotation rate $\Omega$, obtained from one half-cycle of the time-domain measurements shown in Fig.~\ref{Fig:Result_timeDomain} at $f_{\Omega}=0.03$, 0.003, and 0.0006~$\mathrm{Hz}$, respectively.
(d) Response coefficients obtained from 12-h steady-rotation measurements are $k(\beta=1^{\circ})=0.0739\pm0.005$, $k(\beta=10^{\circ})=0.0180\pm0.006$, and $k(\beta=45^{\circ})=0.0059\pm0.0006~(\mathrm{rad/s})^{-1}$. The dashed lines represent linear fits over the selected linear ranges. The response enhancement is evaluated relative to the experimental reference configuration at $\beta=45^{\circ}$, yielding a maximum enhancement of approximately 13.
}
\end{figure}

Figures~\ref{Fig:Result_Response_enhancement}(a)--(c) show representative linear fits obtained by randomly selecting data within one half-cycle of the time-domain measurements in Fig.~\ref{Fig:Result_timeDomain} at $f_{\Omega}=0.03$, 0.003, and 0.0006~Hz, respectively. The extracted slopes exhibit statistical fluctuations that depend on the selected time interval. These fluctuations become more pronounced at lower rotation frequencies. To obtain a more reliable estimate of the response coefficient, we performed steady-rotation measurements over a period of 12~hours for each post-selection angle. The resulting response coefficients are summarized in Fig.~\ref{Fig:Result_Response_enhancement}(d). The steady measurements provide improved statistical precision compared with the single-segment fits obtained from the sinusoidal measurements.

To quantify the experimentally observed response enhancement, we define
\begin{equation}
G(\beta)=\frac{k(\beta)}{k_{\rm ref}},
\label{eq:response_enhancement}
\end{equation}
where $k_{\rm ref}$ is the response coefficient measured in a fixed experimental reference configuration. In the present experiment, the $\beta=45^\circ$ configuration is used as the reference. This reference is defined operationally from the measured detector response under otherwise identical experimental conditions.

As shown in Fig.~\ref{Fig:Result_Response_enhancement}(d), the response coefficient increases systematically as the post-selection angle is reduced. The largest response is obtained at the smallest experimentally investigated angle, $\beta=1^\circ$. Relative to the $\beta=45^\circ$ experimental reference, the maximum measured response enhancement is
\begin{equation}
G_{\rm max}\approx12.5\pm 1.5.
\label{eq:Gmax}
\end{equation}
Here, the uncertainty in the response enhancement is propagated from the fitting uncertainties of the response coefficients $k(\beta=\beta=1^\circ)$ and the reference coefficient $k(\beta=\beta=45^\circ)$.
Thus, the near-orthogonal post-selection condition produces an approximately 13-fold enhancement of the experimentally measured detector response to a given rotation rate.

\subsection{Noise characterization and Allan deviation}

\begin{figure}[t]
\vspace{-0.3cm}
\centering
\includegraphics[width=0.5\textwidth]{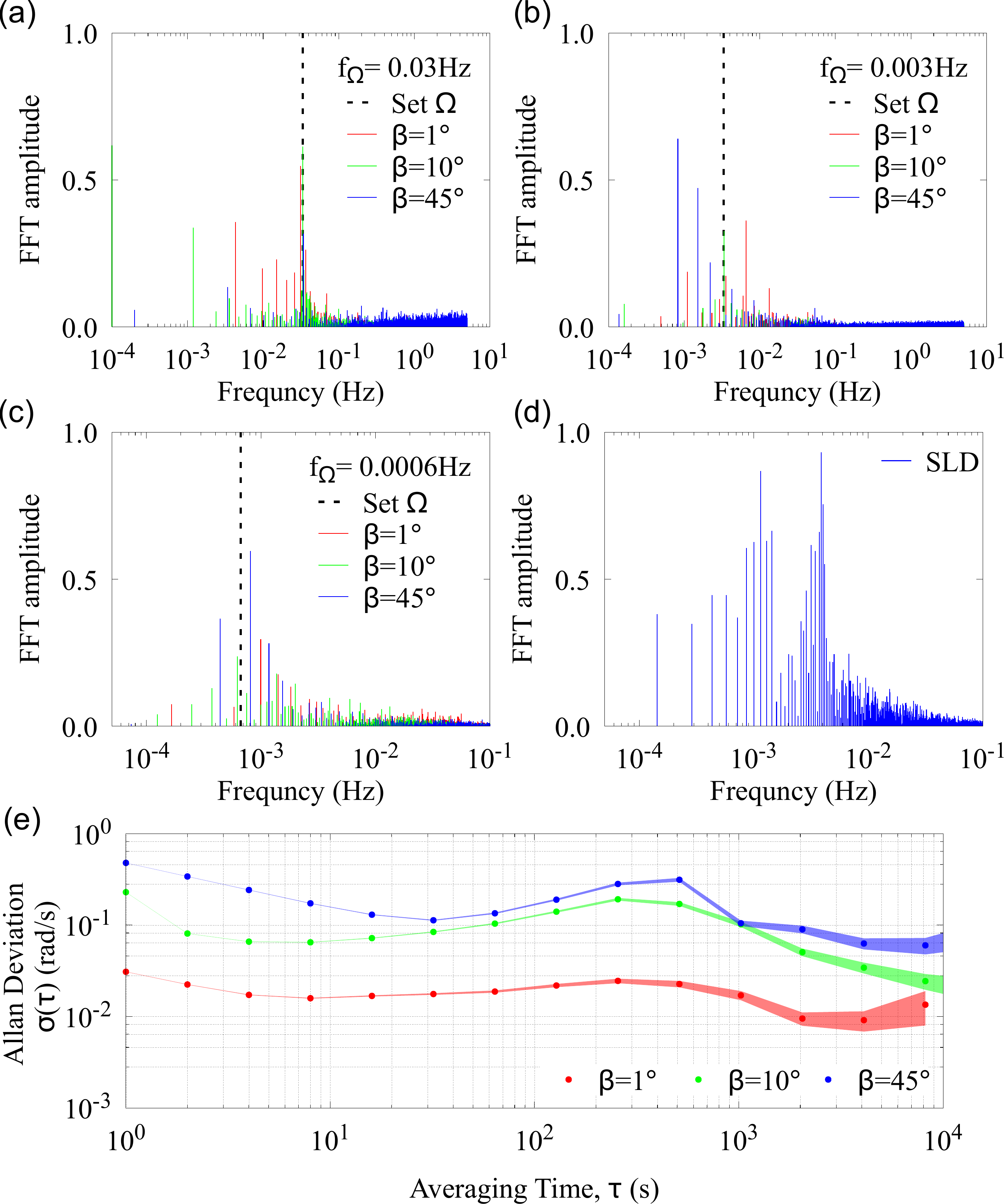}
\caption{\label{Fig:FFT+Allan}
Noise characterization and Allan deviation of the fiber Sagnac gyroscope. (a)--(c) Fast Fourier transform (FFT) spectra of the detected signal $\delta I$ obtained from the time-domain measurements in Fig.~\ref{Fig:Result_timeDomain} for $f_{\Omega}=0.03$, 0.003, and 0.0006~Hz, respectively. (d) FFT spectrum of the output power of the SLD source measured independently under static conditions. (e) Relative Allan deviation of the gyroscope for different post-selection angles. The shaded bands indicate the corresponding $1\sigma$ uncertainty intervals.
}
\end{figure}

Figure~\ref{Fig:FFT+Allan}(a)--(c) shows the fast Fourier transform (FFT) spectra of the time-domain detector signals presented in Fig.~\ref{Fig:Result_timeDomain}. The data were acquired over 2~hours with a sampling frequency of 10~Hz. For the $\beta=45^\circ$ experimental reference configuration, pronounced low-frequency fluctuations are observed in the range of $10^{-4}$--$10^{-2}$~Hz. As the post-selection angle is reduced, the spectral component at the corresponding rotation frequency becomes increasingly pronounced relative to the surrounding background.

To characterize the contribution of the optical source, we independently measured the output power of the SLD under static conditions. As shown in Fig.~\ref{Fig:FFT+Allan}(d), the SLD output was monitored for $7\times10^{3}$~s using a configuration consisting of the SLD, a polarizer, a fiber coupler, a second polarizer, and an avalanche photodiode. Enhanced low-frequency fluctuations are observed in the range of $10^{-3}$--$10^{-2}$~Hz, overlapping with the dominant low-frequency components observed in the gyroscope measurements.

We further evaluated the long-term stability of the gyroscope using static measurements for different post-selection angles. Figure~\ref{Fig:FFT+Allan}(e) presents the relative Allan deviation as a function of averaging time $\tau$ on a log-log scale. The equivalent rotation-rate fluctuation is calculated from the measured detector fluctuations according to
\begin{equation}
\sigma(\tau;\beta)=
\frac{\sigma_{\delta I}(\tau;\beta)}{k(\beta)},
\end{equation}
where $\sigma_{\delta I}(\tau;\beta)$ is the Allan deviation of the measured detector signal and $k(\beta)$ is independently determined from the steady-rotation measurements in Fig~\ref{Fig:Result_Response_enhancement}(d).

%The Allan-deviation curves exhibit regions corresponding to different noise processes. A slope of $-1$ is associated with quantization noise (QN), a nearly zero slope corresponds to bias instability (BI), and a slope of $+1/2$ is characteristic of rate random walk (RRW)~\cite{Allan-Statistics}. The near-orthogonal post-selection configurations exhibit lower equivalent rotation-rate fluctuations than the $\beta=45^\circ$ reference over the corresponding averaging-time ranges.

The Allan-deviation curves exhibit regions corresponding to different noise processes. A slope of $-1$ is associated with quantization noise (QN), a nearly zero slope corresponds to bias instability (BI), and a slope of $+1/2$ is characteristic of rate random walk (RRW)~\cite{Allan-Statistics}. The near-orthogonal post-selection configurations exhibit lower equivalent rotation-rate fluctuations than the $\beta=45^\circ$ reference over the corresponding averaging-time ranges. In particular, the minimum Allan deviation is approximately $0.06~\mathrm{rad/s}$ for the $\beta=45^\circ$ reference configuration, whereas it is reduced to $0.0091~\mathrm{rad/s}$ for the near-orthogonal configuration with $\beta=1^\circ$, corresponding to an improvement by a factor of approximately $6.6$. This reduction demonstrates the enhanced response of the imaginary-WVA configuration in suppressing the equivalent rotation-rate fluctuations at its optimal averaging time.

The Allan-deviation curves do not exhibit a conventional decrease followed by a pronounced increase over the entire averaging-time range. The low-frequency fluctuations observed in the FFT measurements are also reflected in the long-term stability measurements, particularly over averaging times corresponding approximately to $\tau\sim10^{2}$--$10^{3}$~s.

\section{Discussion and Summary}

The experimental results demonstrate that the post-selection condition provides an effective means of controlling the detector response of the fiber Sagnac gyroscope. In particular, the response coefficient $k(\beta)$ increases as the post-selection angle is reduced toward the near-orthogonal condition, reaching a maximum experimental enhancement of approximately 13 relative to the $\beta=45^\circ$ reference configuration. This enhancement is consistent with the expected behavior of imaginary weak-value amplification, in which a small phase perturbation is mapped onto an amplified spectral response through polarization post-selection.

It is important to distinguish the experimentally measured response enhancement from an absolute weak-value amplification factor. In the present work, the response is quantified directly through
$k(\beta)={\partial\delta I}/{\partial\Omega}$
and the relative enhancement is defined as
$G(\beta)={k(\beta)}/{k_{\rm ref}}.$
This approach does not require an independent calibration of an absolute experimental weak value and provides a direct characterization of the experimentally accessible detector response. In particular, the $\beta=45^\circ$ configuration is used only as an experimental reference and is not assigned a finite absolute imaginary weak value.

This distinction is relevant because the ideal theoretical model for the polarization states used in this work and the pioneering work~\cite{PhysRevLett.111.033604} give ${\rm Im}A_w=0$ at $\beta=45^\circ$. Therefore, the measured response of this configuration should not be interpreted as a direct measurement of a nonzero imaginary weak value. A finite experimental response can arise from nonideal effects, including residual fiber birefringence, imperfect polarization compensation, finite post-selection-angle accuracy, polarization-dependent loss, detector imbalance, and other systematic contributions. These effects motivate the use of an operational experimental reference rather than an absolute weak-value calibration.

The experimentally observed maximum enhancement of approximately 13 is lower than enhancement factors reported in some free-space implementations of real- and imaginary-WVA~\cite{PhysRevLett.134.080802,PhysRevLett.111.033604,Huang:26}. 
Such a comparison should be made with caution because the measured enhancement depends not only on the weak value but also on the optical pointer, source bandwidth, spectral detection scheme, polarization control, and detector characteristics. 
Real WVA generally produces a spatial or temporal displacement of the optical pointer, whereas imaginary WVA converts the weak interaction into a spectral displacement that requires a broadband source and spectrally resolved detection~\cite{PhysRevLett.111.033604}. The present implementation combines a polarization-maintaining fiber Sagnac interferometer with a simplified spectral detection system, and therefore its experimentally accessible enhancement is determined by the combined performance of the fiber, polarization-control system, optical source, and detector.

%The distinction between response enhancement and sensitivity is also important. The measured factor $G_{\rm max}\approx13$ directly quantifies the enhancement of the detector response to a given rotation rate. The corresponding equivalent rotation-rate fluctuation is determined by both the detector noise and the response coefficient according to $G(\beta)={k(\beta)}/{k_{\rm ref}}.$ Thus, an increase in $k(\beta)$ improves the conversion of detector fluctuations into equivalent rotation-rate fluctuations only when the accompanying detector noise does not increase by a comparable or larger factor. The Allan-deviation measurements show lower equivalent rotation-rate fluctuations for the near-orthogonal post-selection configurations over the relevant averaging-time ranges, supporting the practical benefit of the enhanced response for the present gyroscope.

The FFT measurements indicate that low-frequency fluctuations of the SLD source constitute an important limitation on the long-term stability. The source-power spectrum exhibits enhanced fluctuations in the $10^{-3}$--$10^{-2}$~Hz range, consistent with the low-frequency components observed in the sinusoidal rotation measurement. These fluctuations also contribute to the non-monotonic behavior of the Allan deviation at long averaging times. Consequently, the present performance is not determined solely by the WVA response but by the combined noise properties of the optical source, polarization system, and detector.

Several improvements could further enhance the performance of the proposed scheme. More accurate polarization control and birefringence compensation would reduce systematic deviations from the ideal polarization model. Higher-resolution spectral detection could improve the readout of the WVA-induced spectral response, while a more stable broadband optical source could suppress the dominant low-frequency noise identified in the present experiment. Alternative pre- and post-selection states may also provide additional flexibility in optimizing the imaginary weak-value response while avoiding a vanishing ${\rm Im}A_w$ at a particular experimental reference angle.

In summary, we experimentally demonstrate a fiber Sagnac gyroscope employing imaginary weak-value amplification. By varying the polarization post-selection angle, we directly measure the response coefficient of the gyroscope and observe a maximum experimental response enhancement of approximately 13 relative to a fixed $\beta=45^\circ$ reference configuration. The enhancement is quantified from the measured detector response and therefore does not require an absolute calibration of the weak value. Noise and Allan-deviation measurements further demonstrate the corresponding improvement in equivalent rotation-rate stability over relevant averaging-time ranges, while identifying low-frequency fluctuations of the broadband SLD source as an important limitation. These results establish imaginary weak-value amplification as a practical approach for enhancing the optical readout response of fiber-based Sagnac gyroscopes and provide a basis for further improvements in polarization control, spectral detection, and source stability.

\section{ACKNOWLEDGMENT}
This study was supported by the National Natural Science Foundation of China (Grant Nos. U25D8008, 42504048, and 42327803). J.-H. Huang acknowledges support from the Hubei Provincial Natural Science Foundation of China (Grant No. 20250650025) and the Fellowship Program of the China National Postdoctoral Program for Innovative Talents (Grant No. BX20250161).

\bibliographystyle{IEEEtran}
\bibliography{references_without_urls}

\newpage

\section{Biography Section}
\begin{IEEEbiographynophoto}{Jing-Hui Huang}
Jing-Hui Huang received the B.S. and M.S. degrees in theoretical physics from the School of Mathematics and Physics, China University of Geosciences (Wuhan), China, and the Ph.D. degree in geophysics from the School of Geophysics and Geomatics, China University of Geosciences (Wuhan), in 2020. He is currently a postdoctoral fellow with the School of Geophysics and Geomatics, China University of Geosciences (Wuhan). From 2022 to 2023, he was a visiting researcher at the University of Ottawa, Canada. His research interests include weak-value amplification, quantum metrology, interferometric sensing, fiber-optic gyroscopes, and precision geophysical sensing.
\end{IEEEbiographynophoto}

\begin{IEEEbiographynophoto}{Hong-Liang Cui}
Hong-Liang Cui is with the Shenzhen Institutes of Advanced Technology, Chinese Academy of Sciences, Shenzhen, China. His research interests include weak-value amplification, optical and fiber-optic sensing, terahertz spectroscopy and imaging, and precision optical measurement. He has contributed to the development of weak-value-amplification-based fiber-optic and terahertz sensing techniques.
\end{IEEEbiographynophoto}

\begin{IEEEbiographynophoto}{Tianying Chang}
Tianying Chang is with the Shenzhen Institutes of Advanced Technology, Chinese Academy of Sciences, Shenzhen, China. His research interests include fiber-optic sensing, interferometric sensing, optical measurement, and terahertz sensing and imaging. His research has included fiber-optic interferometric systems for seismic and tsunami monitoring and weak-value-amplification-based optical sensing.
\end{IEEEbiographynophoto}

\begin{IEEEbiographynophoto}{Jing Zhang}
Jing Zhang received the B.E. degree in optical engineering from Huazhong University of Science and Technology, China, in 2012, the master's degree from Paris-Sud University, France, in 2015, and the Ph.D. degree from Nanyang Technological University, Singapore, in 2019. In 2020, she joined the School of Mechanical Engineering and Electronic Information, China University of Geosciences (Wuhan), China, where she is currently a professor. Her research interests include advanced multifunctional fibers, fiber-shaped optoelectronic devices, optical fiber sensing, and functional fiber devices.
\end{IEEEbiographynophoto}

\begin{IEEEbiographynophoto}{TianYe Huang}
Tianye Huang (Member, IEEE) received the Ph.D. degree in optoelectronic engineering from the Huazhong University of Science and Technology, Wuhan, China, in 2012. From 2010 to 2011, he was at the Photonic Systems Group, McGill University, Canada, as a joint Ph.D. student. From 2013 to 2016, he was with the School of Electrical and Electronic Engineering, Nanyang Technological University, Singapore, as a Research Fellow. Since 2016, he has been a Professor with the School of Mechanical and Electronic Information, China University of Geosciences, Wuhan. His current research interests include integrated optical sensors and nonlinear optical signal processing.
\end{IEEEbiographynophoto}

\begin{IEEEbiographynophoto}{Xiang-Yun Hu}
Xiang-Yun Hu received the Ph.D. degree in mineral resource prospecting and exploration from China University of Geosciences (Wuhan), China, in 2000. He is currently a professor and doctoral supervisor with the School of Geophysics and Geomatics, China University of Geosciences (Wuhan). His research interests include geophysical exploration, electromagnetic methods, deep-resource exploration, weak-signal extraction, and integrated geophysical interpretation. He has published extensively in geophysics and related fields and has led research projects in geophysical exploration and deep-resource detection.
\end{IEEEbiographynophoto}

\vfill

\end{document}